\documentclass[
 reprint,
superscriptaddress,
 amsmath,amssymb,
 aps,
pra,
]{revtex4-2}
\usepackage{amsthm}
\usepackage{mathtools}
\usepackage{physics}
\usepackage{xcolor}
\usepackage{graphicx}
\usepackage{ulem} % Required for \sout
\usepackage[left=23mm,right=13mm,top=35mm,columnsep=15pt]{geometry} 
\usepackage[export]{adjustbox}
\usepackage{placeins}
\usepackage[utf8]{inputenc}
\usepackage[T1]{fontenc}
\usepackage{lipsum}
\usepackage{csquotes}
\usepackage{bm}% bold math
\usepackage{siunitx}
\usepackage{caption}
\usepackage{siunitx}
\usepackage{subcaption}
\usepackage{hyperref}
\usepackage{multirow}
\usepackage{booktabs} 
\usepackage{lineno}
\usepackage{siunitx}
\usepackage{upgreek}

\usepackage[english]{babel}
\usepackage[colorinlistoftodos, color=green!40, prependcaption]{todonotes}

\usepackage{dcolumn}% Align table columns on decimal point
\newcommand{\ox}{Department of Physics, University of Oxford, Parks Road, Oxford OX1 3PU, United Kingdom}

\newcommand{\eq}[1]{Eq. \eqref{#1}}
\newcommand{\fig}[1]{Fig. \ref{#1}}

\begin{document}
% \linenumbers

\preprint{APS/123-QED}
%\title{An X-ray free electron laser based search for heavy axion-like-particles: towards the heavy QCD axion}
\title{Probing keV mass QCD axions with the SACLA X-ray free electron laser}

% \author{Charles Heaton}
% \email[]{charles.heaton@physics.ox.ac.uk}% Your name
% \affiliation{\ox}
% \author{Jack W. D. Halliday}
% \affiliation{Rutherford Appleton Laboratory, Didcot OX11 0QX, UK}
% \author{Taito Osaka}
% \affiliation{RIKEN SPring-8 Center, Sayo-gun, Hyogo, Japan}
% \author{Ichiro Inoue}
% \affiliation{RIKEN SPring-8 Center, Sayo-gun, Hyogo, Japan}
% \author{Sifei Zhang}

% \affiliation{\ox}
% \author{Ahmed Alsulami}
% \affiliation{\ox}
% \author{Joshua T. Y. Chu}
% \affiliation{\ox}
% \author{Mila Fitzgerald}
% \affiliation{\ox}

% \author{Takaki Hatsui}
% \affiliation{RIKEN SPring-8 Center, Sayo-gun, Hyogo, Japan}

% \author{Motoaki Nakatsutsumi}
% \affiliation{European XFEL, Holzkoppel 4, 22869 Schenefeld, Germany}

% \author{Haruki Nishino}
% \affiliation{RIKEN SPring-8 Center, Sayo-gun, Hyogo, Japan}
% \affiliation{Japan Synchrotron Radiation Research Institute, Sayo-gun, Hyogo, Japan}

% \author{Atsushi O. Tokiyasu}
% \affiliation{Researh Center for Accelerator and Radioisotope Science (RARiS), Tohoku University, 1-2-1 Mikamine, Taihaku-ku, Sendai,Miyagi 982-0826, Japan}

% \author{Robert Bingham}
% \affiliation{Rutherford Appleton Laboratory, Didcot OX11 0QX, UK}
% \affiliation{John Anderson Building, University of Strathclyde, Glasgow G4 0NG, United Kingdom}
% \author{Subir Sarkar}
% \affiliation{\ox}
% \author{Gianluca Gregori}
% \affiliation{\ox}

\author{C. Heaton}
\email[]{charles.heaton@physics.ox.ac.uk}% Your name
\affiliation{\ox}
\author{J. W. D. Halliday}
\affiliation{STFC Rutherford Appleton Laboratory, Didcot OX11 0QX, UK}
\author{T. Osaka}
\affiliation{RIKEN SPring-8 Center, Sayo-gun, Hyogo 679-5148, Japan}
\author{I. Inoue}
\affiliation{RIKEN SPring-8 Center, Sayo-gun, Hyogo 679-5148, Japan}
\affiliation{Department of Advanced Materials Science, The University of Tokyo, Chiba 277-8561, Japan}
\author{S. Zhang}

\affiliation{\ox}
\author{A. Alsulami}
\affiliation{\ox}
\author{J. T. Y. Chu}
\affiliation{\ox}
\author{M. Fitzgerald}
\affiliation{\ox}

\author{T. Hatsui}
\affiliation{RIKEN SPring-8 Center, Sayo-gun, Hyogo 679-5148, Japan}

\author{M. Nakatsutsumi}
\affiliation{European XFEL, Holzkoppel 4, 22869 Schenefeld, Germany}

\author{H. Nishino}
\affiliation{RIKEN SPring-8 Center, Sayo-gun, Hyogo 679-5148, Japan}
\affiliation{Japan Synchrotron Radiation Research Institute, Sayo-gun, Hyogo, Japan}

\author{A. O. Tokiyasu}
\affiliation{Researh Center for Accelerator and Radioisotope Science (RARiS), Tohoku University, 1-2-1 Mikamine, Taihaku-ku, Sendai,Miyagi 982-0826, Japan}

\author{R. Bingham}
\affiliation{STFC Rutherford Appleton Laboratory, Didcot OX11 0QX, UK}
\affiliation{John Anderson Building, University of Strathclyde, Glasgow G4 0NG, United Kingdom}
\author{S. Sarkar}
\affiliation{\ox}
\author{G. Gregori}
\affiliation{\ox}

\date{\today} % Leave empty to omit a date

\begin{abstract}

%Our bounds are sufficient to experimentally verify the stability of ALPs for $m_{a } < 0.5$ eV and to comment on the detection of solar ALPs.
\end{abstract}

\maketitle

While the Standard $SU(3)_\text{C} \otimes SU(2)_\text{L} \otimes U(1)_Y$ Model (SM) is phenomenologically very successful as a quantum field theory of the fundamental particles and forces (apart from gravity), it has a serious fine-tuning problem. While charge parity (CP) symmetry is broken by the weak interaction \cite{Christenson:1964fg}, no such violation is seen for the strong interaction \cite{Smith:1957ht,Abel:2020pzs} thus requiring the corresponding parameter in the SM Lagrangian to be unnaturally close to zero: $\theta_\text{QCD}<10^{-10}$.  An elegant solution to this `strong-CP problem' invokes a new `Peccei-Quinn' (PQ) global $U(1)$ symmetry \cite{Peccei:1977hh} which preserves CP invariance in the strong interactions but is spontaneously broken at some high energy scale. This implies the existence of a new pseudo-scalar Nambu-Goldstone boson (akin to the pion) --- named the axion \cite{Weinberg:1977ma,Wilczek:1977pj}. 
% Axions are hypothetical particles, proposed to account for the invariance of CP symmetry in quantum chromodynamics. 
While axions and axion-like-particles (ALPs) are well-motivated by string theory \cite{Svrcek:2006yi} and beyond-Standard-Model extensions \cite{Marsh:2015xka}, they have remained elusive to experimental searches even after significant  effort over many decades. Building on a recent development using an X-ray free electron laser to search for cosmologically favoured axions of mass $m_a \gtrsim 0.01~\text{eV}$, we extend previous bounds on the ALP-photon coupling, $g_{a\gamma\gamma}$, by over an order of magnitude. We exploit the Bormann effect of Laue crystals in a light-shining-through-wall experiment, with broad sensitivity to $m_{a} \lesssim 22 \text{ \rm eV}$. Moreover for $m_{a}\in (3460,3480) \ \text{eV}$ our sensitivity reaches down to the QCD axion coupling prediction, providing the most stringent laboratory constraints in this mass range. 
\paragraph*{}
Extensive efforts are underway worldwide to detect ALPs, in particular the QCD axion in the mass range $10^{-6} \lesssim m_{a} \lesssim 10^{-4} \text{ \rm eV}$ which would provide the observed cosmic dark matter abundance \cite{Preskill:1982cy,Abbott:1982af,Dine:1982ah}. This expectation depends in detail on the high energy extension of the SM which allows for PQ symmetry breaking, e.g. the Kim-Shifman-Vainshtein-Zakarov (KSVZ) model \cite{Kim:1979if,Shifman:1979if} or the Dine-Fischler-Srednicki-Zhitnitsky (DFSZ) model \cite{Dine:1981rt,Zhitnitsky:1980tq}. It also depends on the cosmological history, in particular if the breaking occurred before or after inflation. The case of the `post-inflationary axion' is especially predictive, and attention has focussed on additional contributions to the relic axion abundance in this scenario, e.g. from the decay of topological defects such as cosmic strings that form in the axion field \cite{Gorghetto:2020qws,Buschmann:2021sdq}. There can be further contributions from the decay of axion domain walls \cite{Hiramatsu:2012sc,Ringwald:2015dsf} and recent work has shown that the preferred mass for DFSZ axions to be dark matter is then required to exceed $10^{-2} \ \text{eV}$ \cite{Beyer:2022ywc}. This mass range has been little explored however and this motivates us to search using an X-ray laser for such ``heavy'' axions. 

\paragraph*{}
Most ALP searches utilise their coupling to photons, for which the Lagrangian density is:
\begin{align}
\label{primakoff}
    \mathcal{L} = -\frac{g_{a\gamma\gamma}}{4}\ \mathcal{F_{\mu\nu}}\widetilde{\mathcal{F}}^{\mu\nu}=g_{a\gamma\gamma}\ \mathbf{E}\cdot\mathbf{B}\ a
\end{align}
where $a$ is the ALP field, $g_{a\gamma\gamma}$ is the ALP-photon coupling, and $\mathcal{F_{\mu\nu}}$ is the electromagnetic field tensor. Due to this interaction, detectable photons can be generated from ALPs in a sufficiently strong electric or magnetic field by the Primakoff effect \cite{Primakoff:1951iae}. For clarity, we refer to any pseudo-scalar particle that interacts via \eq{primakoff} as an ALP, and consider the QCD axion to be that class of ALPs for which $g_{a\gamma\gamma }$ is related to the scale of PQ symmetry breaking as \cite{ParticleDataGroup:2024cfk}:
\begin{align}
\label{QCDAxion}
    g_{a\gamma\gamma} = 2\times10^{-16}\ C_{a\gamma}\frac{m_{a}}{\mu \text{\rm eV}} \ \text{\rm GeV}^{-1}
\end{align}
where $|C_{a\gamma}|$ is of $\mathcal{O}(1)$ for both KSVZ and DFSZ models. 

\paragraph*{}
Historically, most searches for ALPs have utilised strong magnetic fields to convert ALPs produced in the Sun as well as other astrophysical sources, or those which make up the dark matter, into detectable photons. These searches can be categorised as helioscopes, which aim to detect ALPs produced in the Sun \cite{CAST:2017uph,Armengaud:2014gea}, or haloscopes, which aim to detect axions that constitute the local Galactic halo dark matter \cite{Brubaker:2016ktl,CAPP:2024dtx,ADMX:2024xbv}. Such searches are mainly aimed towards ALPs lighter than $10^{-4}\ \text{eV}$, but 
notably the CAST helioscope has provided bounds on $g_{a\gamma\gamma}$ extending up to $m_{a}\lesssim 10^{-1}\  \text{eV}$ \cite{CAST:2017uph}. Bounds on axions  can also be derived by considering how the inclusion of axion physics would alter astrophysical observations. This has provided constraints  on $g_{a\gamma\gamma}$ for a wide range of ALP mass from consideration of observations of solar neutrinos \cite{Gondolo:2008dd}, globular clusters \cite{Dolan:2022kul}, supernova neutrinos \cite{Hoof:2022xbe} and the cosmic microwave background \cite{Caloni:2022uya}. It is clear, however, that for both haloscopes and helioscopes and, in particular, for astrophysical bounds, the conclusions are  subject to astrophysical uncertainties. There is thus a specific need for laboratory-based searches as well \cite{Jaeckel:2006xm}. 

\paragraph*{}
In such searches the ALP is both produced and detected in the same experiment, removing any model dependence in the inferred bounds. Leading constraints for heavy axions have been set by experiments at accelerators, such as NOMAD \cite{NOMAD:2000usb}, BABAR \cite{BaBar:2017tiz,Dolan:2017osp}, and NA64 \cite{NA64:2020qwq}. While laboratory-based searches for ALPs cannot yet surpass the sensitivity of astrophysical arguments, they can validate existing bounds from astrophysical observations and give credence towards the models on which they are based.
\begin{figure*}[t!]
    \centering
    \includegraphics[width=\linewidth]{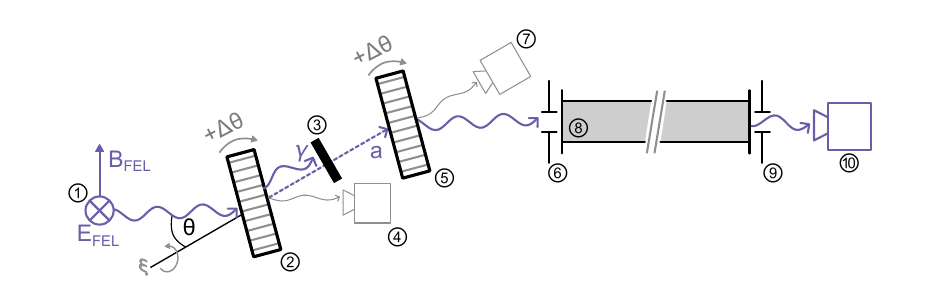}
    \caption{Our experimental setup: $\sigma$ polarised X-rays enter the hutch, \textbf{1}. X-rays impinge on the first Ge crystal. The crystal is mounted on motorised stages to have $\theta$ and $\xi$ rotation control, \textbf{2}. A thick retractable blocker stops the X-rays, but ALPs pass through the blocker, \textbf{3}. To verify alignment, transmission from the first crystal is recorded on the Transmission MPCCD, \textbf{4}. For aid in alignment and to characterise the setup, a second Diffraction MPCCD is placed above the setup, \textbf{7}. The ALPs reach the second Ge crystal where there is a probability, $P_{a\leftrightarrow\gamma}$, that they are reconverted to 10 keV X-rays, \textbf{5}. The second Ge crystal too has $\theta$ and $\xi$ control and can be retracted to record diffraction from the first crystal on the Diffraction MPCCD. We slit the X-rays before entering and after exiting the 2.6 m flight tube, \textbf{8}, with 4 jaw slits, \textbf{6}, \textbf{9}. X-rays are detected on the CITIUS or MPCCD detector, \textbf{10}.}
    \label{fig:setup}
\end{figure*}
Building on previous laboratory experiments, we present results from a new search for heavy axions performed at the EH4c endstation of theSPring-8 Angstrom Compact free-electron LAser (SACLA) \cite{ishikawa2012compact}. Our bounds are applicable to a broad range of ALP masses, $m_{a} \lesssim 22 \text{ \rm eV}$ and also to for keV mass ALPs, $m_{a}\in (3460,3480)$~eV. 
 
\section*{THEORY \& EXPERIMENT}

Our experiment employs the Primakoff effect to convert X-ray photons
into ALPs via their interaction %of the magnetic field of the X-ray photons 
with the static electric field of the crystal lattice under Laue's diffraction conditions. Both the diffracted X-rays and generated ALPs then impinge on a wall that blocks the X-rays but allows the feebly interacting ALPs to go through. The ALPs then enter a second crystal and are converted by the inverse Primakoff effect back into X-ray photons which we detect. This light-shining-through-walls (LSW) approach was %proposed by Buchmüller and Hoogeveen 
first discussed for both Bragg and Laue geometries \cite{Buchmuller:1989rb}, and developed further %by Liao 
for Bragg-diffracted X-rays \cite{Liao:2010ig}. Subsequently, %Yamaji et al. 
the advantages of the Laue-crystal approach was emphasised  \cite{Yamaji:2017pep}; the effective path length of X-rays through the crystal, $L_{\text{\rm eff}}$, being then enhanced by the anomalously high transmission and diffraction efficiency due to the Borrmann effect \cite{Kovev1969,borrmann1950absorption}.

% This "light-shining-through-walls" approach has been carried out previously but with magnetic fields and optical lasers \cite{Robilliard:2007bq,Ehret:2010mh,OSQAR:2015qdv} for bounds for $m_{a} < 2\times10^{-4} \text{ \rm eV}$. 
\paragraph*{}
Such X-ray LSW experiments using Laue-crystals have  been carried out at a synchrotron \cite{Yamaji:2018ufo} and a X-ray free electron laser (XFEL) \cite{Halliday:2024lca}. The exceedingly high electric field between crystal planes in Ge crystals, $E_{\text{\rm eff}}=7.3 \times 10^{10} \ \text{Vm}^{-1}$, is equivalent to a magnetic field of $\sim 1 \ \text{T}$ -- stronger than typical electromagnets. Hence despite the much shorter path length, the integrated effect is comparable to what is achievable in  optical LSW experiments. The sensitivity to ALP mass is expressed via momentum conservation by the relation \cite{Yamaji:2017pep}
\begin{align}
\label{eq:m_a}
    \left|m_{a}^{2}- m_{\gamma}^{2} - 2q_{T}\left[k_{\gamma}\sin\left(\theta_{B} +\Delta \theta\right) - \frac{q_{T}}{2}\right]\right| <\frac{4k_{\gamma}}{L_{\text{\rm eff}}},
\end{align}
where $m_{\gamma}$ is the plasma frequency in the crystal, $q_{T}$ is the reciprocal lattice vector, and $k_{\gamma}$ is the X-ray photon energy. (We use Heavyside-Lorentz units, $\hbar=c=1$, throughout unless otherwise stated.)

\paragraph*{}
Our setup shown in \fig{fig:setup}  consists of two Ge(220) crystals of thickness $\ell = 1.5$ mm, with a movable opaque block between them. SACLA beamline 3 \cite{tono2013beamline} was operated at 30 Hz in self-seeded mode \cite{inoue2019generation} at $k_{\gamma} = 10 \text{\ \rm keV}$. A Si (111) double channel-cut crystal monochromator was placed upstream of the hutch to cut the SASE pedestal around $\Delta E \approx 1.3\   \text{\rm eV}$ ($\Delta E/E\sim10^4$) \cite{katayama2019x}. The Laue crystals were rotated such that the effective electric field between the (220) planes, $\mathbf{E}_{\text{\rm eff}}$, is oblique to the magnetic field $\mathbf{B}_{\rm X}$ of the incident X-rays, hence:
\begin{align}
\mathcal{L}=g_{a\gamma\gamma}\ |\mathbf{E}_{\text{\rm eff}}||\mathbf{B}_{\text{\rm X}}|\sin\left(\theta_{B}\right) a,
\end{align} where $\theta_{B}$ is the Bragg angle. To align the crystals, two multi-port charge-coupled device (MPCCD) detectors \cite{kameshima2014development} were placed at $2\theta_{\text{\rm B}}$ to record transmission and diffraction effective rocking curves of the first crystal. Once aligned, the second crystal was driven into position and aligned onto a third MPCCD downstream. Both crystals were mounted onto motorised rotation and swivel stages to allow for adjustments in the $\theta$ and $\xi$ directions, shown in \fig{fig:setup}. The crystals used in our experiment were cut from a larger monolithic crystal. While Laue crystals are ideally cut such that planes are perpendicular with the crystal surface, this cut is inevitably not perfect and so a small (here measured to be on the order of $1^{\circ}$) angular misalignment needs to be corrected for with a $\xi$ rotation.
%\todo[]{You need to define more precisely what you mean by miscut.} 
Once aligned, the downstream MPCCD was driven out and replaced with the more sensitive CITIUS detector \cite{ozaki202317400}, which was used for the axion searches.

\paragraph*{}
Axions are produced coherently when X-rays impinge at an angle $\theta = \theta_{B}+\Delta \theta$ on a Laue crystal. In principle, the angular width of this resonance is set by $4k_{\gamma}/L_{\text{\rm eff}}$, where 
$L_{\text{\rm eff}}$  is the effective path length of the photons through the crystal. In practice, however, the resonance width is dominated by the bandwidth of the XFEL, which therefore determines the experimental angular acceptance. By detuning the crystals from their aligned Bragg angles, sensitivity is obtained to heavier axions with mass $m_{a}$ which solves Eq.~\eqref{eq:m_a}. %These generated ALPs will exit the crystal at a angle of $\theta_{B}-\Delta \theta$, or $2\theta_{B}$ from the original path of the X-rays. One expects that the ALPs will be detected at the same position as the double diffracted X-rays but, for $\Delta \theta > 0$, with a small vertical offset due to the ALPs being channelled at a different angle
%\begin{align}
%\label{deltaZ}
%    \Delta z = l\left(\frac{1}{\sin\left(\theta_{\text{\rm B}} + \Delta \theta\right)} - \frac{1}{\sin\left(\theta_{\text{\rm B}}\right)}\right).
%\end{align}
For $\Delta \theta = 0$, this happens when \cite{Liao:2010ig,Yamaji:2017pep}:
\begin{align}
%    \sin\left(\theta_{B} \pm \frac{\Delta \theta^{\text{\rm eff}}_{\text{\rm RC}}}{2}\right)=\frac{q_{T}}{2 k_{\gamma}}+\frac{m_{a}^{2}-m_{\gamma}^{2}}{2q_{T}k_{\gamma}},
    \sin\left(\theta_{\text{\rm B}}\right)=\frac{q_{T}}{2 k_{\gamma}}+\frac{m_{a}^{2}-m_{\gamma}^{2} + \Delta m_{\gamma}^{2}}{2q_{T}k_{\gamma}},
\end{align}
where $\Delta m_{\gamma}$ is a change in $m_{\gamma}$ as a result of the standing wave in the crystal defined as $\Delta m_{\gamma}=m_{\gamma}\sqrt{f(q_T)/f(0)}$ and $f$ is the real part of the atomic form factor. Given $2q_{T}k_{\gamma} \gg |m_{a}^{2}-m_{\gamma}^{2} +\Delta m_{\gamma}^{2}|$, this condition is met for any $m_{a} \lesssim \sqrt{m_{\gamma}^{2}-\Delta m_{\gamma}^{2}}$ when $\theta\approx\theta_B$. %and ALPs are indiscriminately generated for $m_{a} \lesssim m_{\gamma}$ in X-ray / crystal ALP generation experiments.
\begin{figure}[b!]
    \centering
\includegraphics[width=\linewidth]{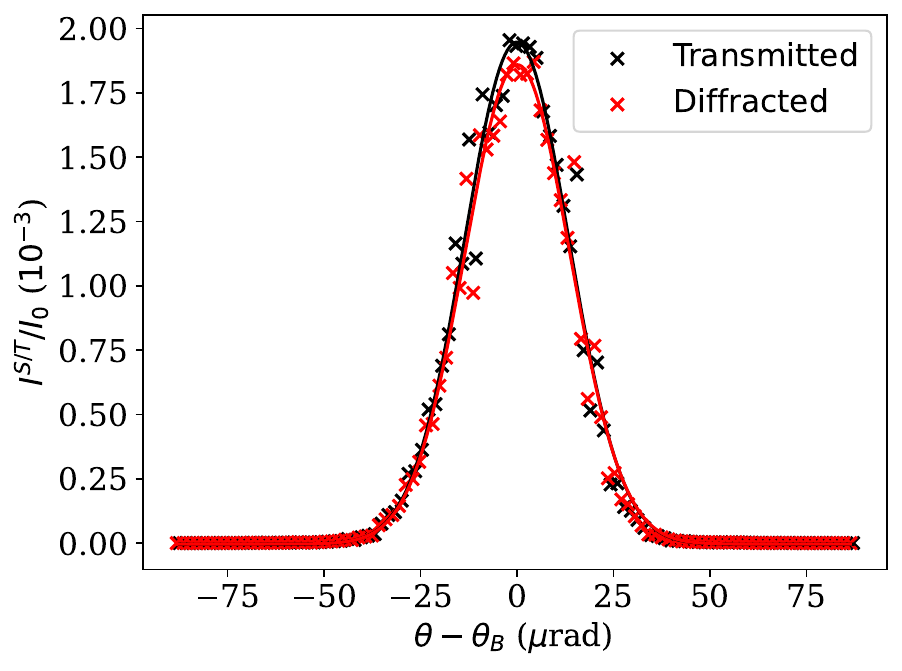}
    \caption{Diffracted and transmitted efficiencies of the first Ge crystal; as expected for Laue diffraction: $I_{\text{\rm Ge}}^{S} \approx  I_{\text{\rm Ge}}^{T}$. Both measurements are corrected for attenuation due to scattering in air \cite{Henke:1993eda}. The fit to the  rocking curve has width $33$ $\upmu$rad and peak diffracted efficiency of $1.9\times10^{-3}$.}
    \label{fig:I_ST}
\end{figure}
\paragraph*{}
For both ``detuned'' ($\Delta \theta \neq 0$) and ``on-Bragg'' ($\Delta \theta = 0$) searches, the probability of the conversion of an X-ray to an ALP (and vice versa) is given by \cite{Yamaji:2017pep}
\begin{align}
\label{eq:probability}
    P_{a\leftrightarrow\gamma} = \left[\frac{1}{2}g_{a\gamma\gamma}E_{\text{\rm eff}}L_{\text{\rm eff}}\nu_{\text{\rm B}}D\cos{\left(\theta_{B} + \Delta \theta\right)}\right]^{2},
\end{align}
where $D$ is defined as
\begin{align}
    D=
    \begin{cases}
       &D_{a\rightarrow \gamma} = \frac{k_{a}}{q_{T}} \frac{\sin\left(2\theta_{\text{\rm B}}\right)}{\cos\left(\theta_{\text{\rm B}} - \Delta\theta\right)}\  \ \   \text{for $\gamma \rightarrow a$},\\
       &D_{\gamma\rightarrow a} = \frac{k_{a}^{2}}{k_{\gamma}q_{T}} \frac{\sin\left(2\theta_{\text{\rm B}}\right)}{\cos\left(\theta_{\text{\rm B}} - \Delta\theta\right)}\ \text{for $a \rightarrow \gamma$},
    \end{cases}
\end{align}
with $k_a$ the axion energy.
For the range of detuning angles considered in this experiment, $D\cos(\theta_{\text{\rm B}} + \Delta \theta) \sim \cos(\theta_{\text{\rm B}})$. 
Because of the
Borrmann effect, we see the formation of two distinct Bloch waves. One with antinodes on the lattice planes and the other with antinodes within planes. The former is strongly absorbed, while the latter propagates with minimal absorption. Since both Bloch waves carry the same amount of energy, and under the Borrmann condition, one expects equal intensities of diffracted and transmitted X-rays, we only consider the quarter of X-rays that diffract efficiently to reach our conversion crystal. This implies an effective probability $P^{*}_{a\leftrightarrow\gamma} = P_{a\leftrightarrow\gamma}/4$.
%\todo[]{Note: this is a very important point.}
%results in a suppression of $P_{a\leftrightarrow\gamma}$ for $m_a \lesssim m_{\gamma}$ by a factor of 4. %at $m_{a} = m_{\gamma} \pm \Delta m_{\gamma}$, where $\Delta m_{\gamma} :=m_{\gamma}\sqrt{\frac{f(q)}{f(0)}}$ where $f(q)$ is the atomic form factor evaluated at $q=\frac{4\pi}{\lambda} \sin\left(\frac{\theta_{\text{\rm B}}}{2}\right)$.
\eq{eq:probability} differs from previous synchrotron based ALP searches with the inclusion of $\nu_{\text{\rm B}}$ \cite{Yamaji:2017pep,Yamaji:2018ufo,Halliday:2024lca}, where $\nu_{\text{\rm B}}$ reflects the fact that the rocking curve becomes much narrower when the X-ray pulse is sufficiently short \cite{wark,Shvydko:2012rzc}. This is the case in the present work with pulses delivered at SACLA being on the order of a few to ten fs \cite{yabashi2015overview}. When accounting for dynamical diffraction, the rocking-curve width is set by the time-bandwidth product and can therefore be estimated as \cite{Halliday:2024lca}
\begin{align}
   \Delta \theta_{\text{\rm RC}} \approx 
   \frac{\lambda_{\text{\rm X}}}{2 \ell \sin{\left(\theta_{\text{\rm B}}\right)}} 
   % = \frac{1.24\text{ \rm \AA}}{2\ \times\ 1.5 \text{\ \rm mm} \sin{\left(18.0 \text{\ \rm deg}\right)}} 
  = 0.134 \ {\rm \upmu rad},
\end{align}
for $\ell=1500 \ \upmu\text{\rm m}$ and $\lambda_{\text{\rm X}} = 1.24 \text{\ \rm \AA}$. The parameter $\nu_{\text{\rm B}}$ is defined as the ratio of the crystal's effective rocking curve width to its Darwin width \cite{xrayDatabase}
\begin{align}
\nu_{\text{\rm B}} = \frac{\Delta\theta_{\text{\rm D}}}{\Delta\theta_{\text{\rm RC}}} = \frac{43.4\  \upmu\text{\rm rad}}{0.134 \ \upmu \text{\rm rad}} = 324.  
\end{align}
Here we define the rocking curve width, $\Delta\theta_{RC}$ at the {\it actual} angular acceptance of the crystal, whereas the Darwin width is the {\it ideal} width, neglecting time dependence. 

\paragraph*{}
To maximise $L_{\text{\rm eff}}$ through our double crystal setup, we choose the $\ell$ to be close to the Borrmann length for the standing wave that has its antinodes between planes:
\begin{align}
    L_{\text{\rm eff}} = 2L_{\text{\rm B}}^{\text{\rm att}}\left[1-\exp\left(1-\frac{L_{y}}{2L_{\text{\rm B}}^{\text{\rm att}}}\right)\right],
\end{align}
where for $\sigma$ polarised X-rays, $L_{\text{\rm B}}^{\text{\rm att}} = 1598 \ \upmu \text{\rm m}$ \cite{xrayDatabase} and $L_{y} = \ell/\sin{\left(\theta_{\text{\rm B}} + \Delta \theta\right)}$.

\paragraph*{}
As well as increasing the axion conversion probability, a thicker crystal helps to mitigate the heating of the first Ge crystal, which was found to be problematic during a previous axion search at an XFEL \cite{Halliday:2024lca}. To monitor this, a MPCCD detector is placed in transmission behind the first crystal. When the crystal is aligned to its Bragg angle, the Borrmann effect reduces absorption by a factor of $\sim 30$ \cite{Henke:1993eda,xrayDatabase}. If the first crystal is heated sufficiently to distort crystal planes, we would see a loss of transmitted signal as a result of the loss of the Borrmann effect.  To limit the heating of the first crystal, we defocused the X-ray beam to a spot of $\sim 500 \ \upmu\text{\rm m}$, kept the two Ge crystals in air, and used metal holders to carry heat away from the crystals. The defocused beam also increases diffracted efficiency. The choice to run the experiment in air inevitably increases the background due to X-ray scattering in air. Furthermore, to limit stress induced on the crystal, they are secured adhesively in their holders with a paraffin wax.

\section*{Results}
\fig{fig:I_ST} shows transmission, $I_{\text{\rm Ge}}^{T}/I_{0}$, and diffraction efficiencies, $I_{\text{\rm Ge}}^{S}/I_{0}$, from the first Ge Laue crystal. The measured effective rocking curve width is 33 $\upmu$rad, implying an energy bandwidth of $\Delta E = 1.02$ eV (determined the bandwidth of the free electron laser), with a diffracted efficiency of 1.9$\times10^{-3}$. It is interesting to note that this is only a factor of $1.6$ less than in our previous experiment \cite{Halliday:2024lca} despite the crystals in the present work being 3 times as thick. This is a direct consequence of the Borrmann effect, allowing for the Bloch wave with antinodes in between planes to have an anomalously high transmission through the crystal. Also, from \fig{fig:I_ST}, the diffracted and transmitted intensities are nearly equal, which agrees with scattering occurring under Borrmann conditions.
%This scaling does not line up with Persson's measurements, taken with  and may reflect an interesting aspect of dynamic diffraction with femtosecond pulses \cite{persson1971laue}. 
\begin{figure}[h!]
    \centering    \includegraphics[width=0.8\linewidth]{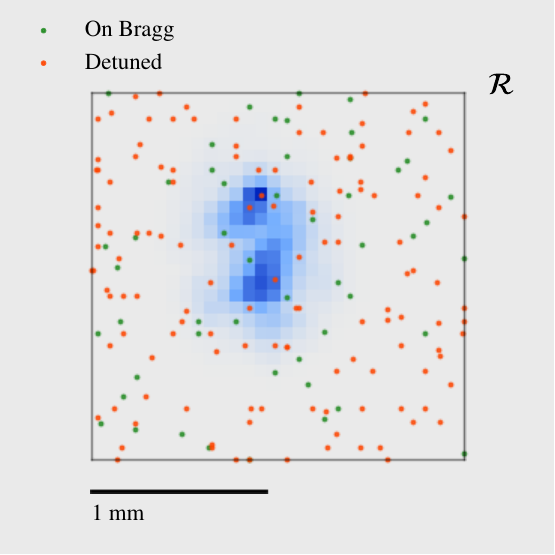}
    \caption{A hit-map of detected photons across all of the axion search data plotted over the intensity profile of the double diffracted X-rays, in blue. A droplet algorithm %\cite{hruszkewycz2012high}% 
    is applied to the frames of CITIUS data to attribute single photon events. The region $\mathcal{R}$ encloses the photon hits.}
    
    %\todo[inline]{This figure needs some changes. The X-ray intensity profile due to the double crystal diffraction should be plotted with a differen color (blue? green?), otherwise they can be misinterpreted as "axion" hits.}
    
    \label{fig:allSearches}
\end{figure}

Once both crystals were aligned, the efficiency factor, $\eta$ was calculated by measuring the flux on the CITIUS detector after the X-ray pulse is diffracted by the two crystals, $I^{\text{\rm CITIUS}}$; this accounts for losses due to scattering in air and imperfect parallelism between the two crystals
\begin{align}
    \eta = \left(\frac{I^{S}_{\text{\rm Ge}}}{I_{0}}\right)^{-2} \frac{I^{\text{\rm CITIUS}}}{I_{0}} \sim 0.629,
\end{align}
where $I_{\text{\rm Ge}}^{\text{\rm S}}$ is the flux of diffracted X-rays, \fig{fig:I_ST}, and $I_{0}$ is the flux of incoming X-rays.
This value is attributable to  scattering in air -- the X-rays travel through $\sim$ 75 cm of air which would attenuate a 10 keV beam by a factor of 0.648 \cite{Henke:1993eda} -- indicating that the crystals are aligned such that their Bragg acceptances overlap within the XFEL bandwidth. 
\paragraph*{}
The rotation stages used for the detuning of the crystals were calibrated by recording the motor positions at (220), ($\Bar{2}$$\Bar{2}$0) and (440) reflections of the Ge crystals. Using these reflections, and assuming $2d_{(220)}=4.000 $ \AA, 
as specified by the manufacturer, we calibrate the stage for the first crystal to step 0.10016 mdeg/pulse with an uncertainty of 2.73 $\times 10^{-6}$ mdeg/pulse and the second crystal's stage steps 0.20012 mdeg/pulse with an uncertainty of 5.29 $\times 10^{-6}$ mdeg/pulse. Assuming uncertainties in the stage positioning are linear with the number of pulses, we estimate the uncertainty in positioning when detuning the crystals to 6.5 deg is $\sim 3.10\ \upmu$rad and 6.00 $\upmu$rad for the first and second crystals respectively. 
\paragraph*{}
Any uncertainty on the axion mass is set by the accuracy of the first crystal's rotation. The axion mass that satisfies \eq{eq:m_a} becomes less constrained as there are now a possible range of $\Delta \theta$ within the rotation stage's uncertainty in position. However, since the error at this detuning angle is below the width of the effective rocking curve, we do not account for additional uncertainty in the axion mass beyond the effect of the XFEL bandwidth. During our on-Bragg searches, transmission from the first crystal was recorded to monitor heating and subsequent loss of alignment. During the 11 hour data collection, we did not see any loss of the transmitted signal and so conclude that photoabsorption was not enough to heat the first crystal sufficiently to lose alignment, see Supplementary Information. We progressed with the detuned scan under the assumption that the crystals would not fall out of alignment due to heating and noted that transmission was always maintained when crystals were rotated back to the Bragg angle. 

\paragraph*{}
Our ALP search data comprises of an on-Bragg search of $\sim 11$ hours and 8 detuned searches at different ALP masses, ranging from $\sim 2 - 7$ hours of collection, summarised in Table \ref{Results}. 
\begin{table*}[ht]
\centering
\begin{tabular}{|c|c|c|c|c|c|c|c|c|}
\toprule
$\Delta \theta$ (deg) & $m_{a}$ (eV) & $N_{\text{\rm in}} \ (\times 10^{15})$ & $N^{(90)}_{\text{\rm det}}$ & $N^{(90)}_{\text{\rm det}}/N_{\text{\rm in}} \ (\times10^{-16})$ & $\ln\left(\text{\rm BF}\right)$ & $g_{a\gamma\gamma}$ $(\times 10^{-6}$ GeV$^{-1}$)  \\
\midrule
0 & $<22$ & 43.7 & 13.5 & 3.09  & -1.89 & 6.50\\
6.5 & 3482 & 27.1 & 9.65 & 3.56 & -2.63& 6.81\\
6.49 & 3479 & 21.4 & 9.62& 4.50 & -1.78& 7.22\\
6.48 & 3476 & 13.6 & 10.8 & 7.96 & -1.31& 8.33\\
6.47 & 3473 & 8.94 & 10.1 & 11.3 & -0.83& 9.10\\
6.46 & 3470 & 7.50 & 7.07 & 9.43 &-2.08 & 8.69\\
6.45 & 3467 & 7.35 & 6.63  & 9.02&-2.12 & 8.59\\
6.44 & 3464 & 7.83 & 6.92 & 8.84  &  -2.47& 8.55\\
6.43 & 3461 & 7.70 & 5.60  & 7.27& -1.11 & 8.14\\
\bottomrule
\end{tabular}
\caption{Summary of the data that forms the axion bounds from our XFEL light shining through a wall experiment. For all detuning angles, BF$<1$ implying a lack of support for the ALP model. $N_{\text det}^{(90)}$ is the fitted upper limit on a 90\% confidence interval on the ALP signal contribution.}

%$N_{\text{\rm det}}$ is defined as the number of photons within the CITIUS direct exposure region of interest that are measured to be $k_{\gamma} \in (8.67,10.8)$ keV as illustrated in \fig{fig:allSearches}. }
% \todo[inline]{CH: Include Bayes factor for each run. Can help to show that axions were not found for any run. Maybe get rid of $L_{eff}$ given it doesn't change that much.}
% \todo[inline]{For on Bragg is ma<44 eV or ma<58 eV as stated elsewhere in the manuscript. Please correct. Also, I would probably add another column to show the value of Leff, as well as the bound on $g_{a\gamma\gamma}$.}
\label{Results}
\end{table*}

While \SI{10}{\kilo\electronvolt} photons are detected on the CITIUS detector with the X-ray blocker in place, we do not attribute this signal to ALP conversion. The data show that the detected photons are spatially confined to a $2 \times \SI{2}{\milli\metre\squared}$ square aperture, as shown in Fig.~\ref{fig:allSearches}. This localisation is consistent with X-ray scattering in air downstream of the double-crystal apparatus.

The scattered radiation is spatially defined by slits positioned both upstream and downstream of the flight tube (see Fig.~\ref{fig:setup}), which were set to a width of \SI{2}{\milli\metre} in both the horizontal and vertical directions. As a result, an otherwise uniform scattered background is confined to a region surrounding the transmitted X-ray beam.
In contrast to our previous experiment performed under vacuum conditions \cite{Halliday:2024lca}, operation in air inevitably produces a higher background due to X-ray scattering.

\begin{figure}[h]
\includegraphics[width=\linewidth]{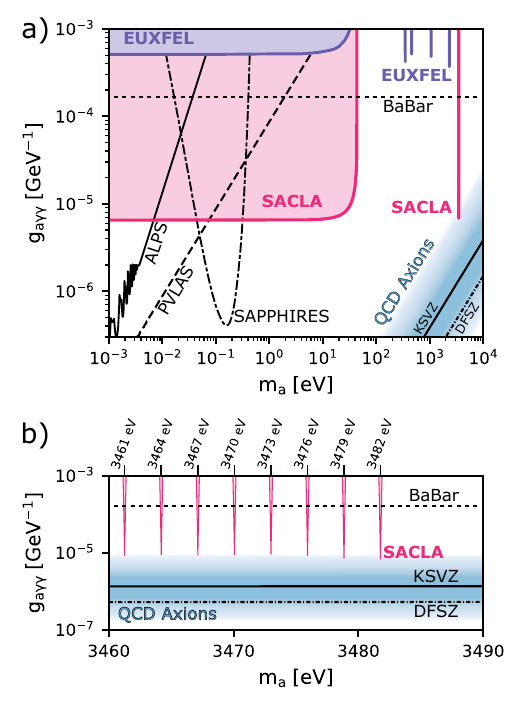}
\caption{Bounds on $g_{a\gamma\gamma}$ (90\% C.L.) derived from our X-ray-shining-through-walls experiment at SACLA represent a significant improvement over the previous such experiment at EuXFEL \cite{Halliday:2024lca}. While a broad mass region is excluded for $m_{a} <22 $ eV, for $m_{a} \in (3462,3482)$ our sensitivity reaches down to the QCD coupling (blue band) obtained from \eq{QCDAxion} 
%for $\left|5/3-1.92\right|<\left|C_{a\gamma}\right| < \left|44/3-1.92\right|$ 
\cite{ParticleDataGroup:2024cfk}. The width of the bounds is set by the width of the effective rocking curve of the Ge crystals. Other experimental bounds \cite{AxionLimits} are shown for comparison: NOMAD \cite{NOMAD:2000usb}, BaBar \cite{BaBar:2017tiz}, SAPPHIRES \cite{Nobuhiro:2020fub,SAPPHIRES:2022bqg}, ALPS \cite{Ehret:2010mh} and PVLAS \cite{DellaValle:2014wea}.}
\label{fig:bounds}
\end{figure}

\paragraph*{}
Despite this experimental consistency, we consider the possibility that the detected photons arise from interactions involving ALPs by evaluating Bayes factors ($\mathrm{BF}$), i.e.,\ by comparing the marginal likelihoods of two competing hypotheses \cite{Trotta:2008qt}. If the detected photons arose from axion production, they would be expected to follow the same spatial distribution as the double-diffracted X-rays (Fig.~\ref{fig:allSearches}), with an additional contribution from a uniform background. Our null hypothesis is that the detected photons arise exclusively from X-ray scattering in air, for which we expect a uniform spatial distribution across the region bounded by the slits. We therefore compute the ratio of the marginal likelihoods for a mixed normal--uniform distribution and for a purely uniform distribution.
For all detuning angles, we do not observe support for the detection of ALPs as BF$ \ll 1$ for all axion masses, Table I, and expect detected photons form a uniform background due to scattering in air. Details on the computation of BF can be found in the Supplementary Information.
%However, we can still achieve competitive bounds from this measurement. 

% To form our bounds , we consider the number of $\sim$10 keV photons that fall within the elliptical region of interest in \fig{fig:allSearches}b. %From \eq{deltaZ}, $\Delta z \approx 158$ $\mu$m which is much smaller than the deviation in the X-ray spot's position and so the same region of interest is used for both on Bragg and detuned data.
To infer bounds on $g_{a\gamma\gamma}$, we take $\left(P^{*}_{a\leftrightarrow\gamma}\right)^{2} = N_{\text{\rm det}}/\eta N_{\text{\rm in}}$, where $N_{\text{\rm det}}$ is the ALP signal contribution to count on the detector (see Table \ref{Results}) and $N_{\text{\rm in}}$ is the total number of photons incident on the first crystal, to find
\begin{align}
    g_{a\gamma\gamma} \leq \left[\frac{1}{4}E_{\text{\rm eff}}L_{\text{\rm eff} }\nu_{\text{\rm B}}\cos\left(\theta_{\text{\rm B}} + \Delta \theta \right)D\right]^{-1}\left(\frac{N_{\text{\rm det}}}{\eta N_{\text{\rm in}}}\right)^{\frac{1}{4}},
\end{align}
for each detuning angle. To obtain a $90\%$ confidence level bound, we fit the spatial distribution of hits to the mixed model and find the 90$\%$ C.L. upper limit of the normally distributed contribution. Although the Bayes-factor analysis described above indicates no evidence for a signal consistent with ALP interactions, we nevertheless set upper limits using this conservative procedure. Our results are compared to other laboratory searches in \fig{fig:bounds}.

\section*{Discussion and outlook}
Our bounds on $g_{a\gamma\gamma}$ for $m_{a} \gtrsim 0.3$ eV are the most stringent obtained from experiment as shown in \fig{fig:bounds}. While for $m_{a}\lesssim 22$ eV our bounds do not reach the QCD axion predictions, they still provide important insights on constraints from astrophysical observations. In particular, high-resolution spectrometry by the James Webb Space Telescope (JWST) has enabled bounds to be derived on $g_{a\gamma\gamma}$ by considering the decay of ALPs to infrared photons ($0.1 \lesssim m_{a} \lesssim 3$ eV) \cite{Janish:2023kvi,Roy:2023omw,Pinetti:2025owq}. Our experimental bounds confirm that for $m_{a} < 0.35$ eV, one should not expect to see photons from spontaneous ALP decay. Since the rate of such decays is \cite{ParticleDataGroup:2024cfk}
\begin{align}
    \Gamma_{a\rightarrow\gamma\gamma} = \frac{g_{a\gamma\gamma}^{2}m_{a}^{3}}{64\pi},
\end{align}
using $g_{a\gamma\gamma} < 6.50 \times10^{-6}$ GeV$^{-1}$, we estimate a lifetime $\tau = \Gamma_{a\rightarrow\gamma\gamma}^{-1} > H_{0}^{-1}$, where $H_{0}$ is the Hubble constant. Hence any detection by JWST of photons with wavelengths $\lambda < 7.08 \ \upmu$m cannot be attributed to ALP decays. 
% \todo[inline]{CH: Edit on 16/02: conversion from energy to photon wavelength wasn't accurate. Doesn't change the conclusion}

\paragraph*{}
Conversely, our bounds for $m_a \sim 3.5$ keV imply an ALP lifetime $\tau > 2.7$ days. Such heavy axions cannot therefore contribute to the overall emission from the Sun as was proposed \cite{vanBibber:1988ge}. 
%It follows that if these heavier ALPs are generated in the Sun, they will not decay to soft X-rays without the presence of a strong, external electromagnetic field or gravitational trapping as $\tau \gg 1 \text{ \rm a.u.}\ /\ c$. 
Thus, while laboratory searches such as ours cannot yet compete in sensitivity with astrophysical arguments, the results obtained are complementary. %gives confidence to the models and background subtraction procedures used to derive astrophysical bounds and, at the same time, guide future searches.

Having demonstrated an experimental sensitivity relevant to QCD axions via favourable scaling of $g_{a\gamma\gamma}$ to the crystal length and establishing a strategy to monitor crystal heating, we discuss a setup suited for investigations of keV mass axions. With measures such as Peltier cooling to mitigate against increased crystal heating, an in-vacuum experiment at a higher X-ray fluence could open up new areas of investigation. One could utilise the 2.25 MHz operation at the European XFEL to maximise X-ray fluence, e.g. assuming background levels comparable to the previous such experiment \cite{Halliday:2024lca} and with a 500 $\mu$J seeded pulse over 65 minutes of acquisition, one can reach the coupling predicted by the DSFZ model for $m_{a} \sim 7.1$ keV. This mass is motivated by the unexplained $\sim 3.5$ keV X-ray emission recorded from the Andromeda galaxy \cite{Boyarsky:2014jta} and galaxy clusters \cite{Bulbul:2014sua} as it was hypothesised that such emission may result from the decay of 7.1 keV mass ALPs \cite{Higaki:2014zua}. There is controversy around whether this line indeed originates from dark matter decay \cite{Dessert:2018qih} and whether such X-ray emission is even present \cite{Dessert:2023fen}. Our experimental platform is well-suited to a laboratory investigation of whether the QCD axion can indeed be responsible for this possible astrophysical X-ray signal.
\section*{Materials \& Methods}
For this experiment, the CITIUS detector had an exposure time of 2 $\mu$s, 28 times shorter than the standard exposure time used at SACLA.
In addition, the CITIUS was operated in a multi-frame sampling mode to reduce readout noise fluctuations and to achieve higher energy resolution. 
Details on the calibration of the incoming photon flux, $I_{0}$, quantum efficiency of the transmission MPCCD, evaluation of crystal heating and the droplet algorithm utilised on the CITIUS data is given in the Supplementary Information.
The Bayes Factor, $BF$, is defined as 
\begin{align}
    \text{\rm BF} = \frac{\int P(D|M_{\text{\rm Mixed}},w) P(w)\text{\rm d}w}{\int P(D|M_{\text{\rm Uniform}},w) P(w)\text{\rm d}w}
\end{align}
where $P(D|M_{\text{\rm Mixed/Uniform}},w)$ is the probability of measuring data, $D$, for the Mixed/Uniform model with parameter $w$ (with prior probability $P(w)$). The two models are defined as
\begin{align}
&P(D|M_{\text{\rm Uniform}}) = \frac{A_{\rm U}}{|\mathcal{R}|} \text{\rm \ \ for } x,y \in \mathcal{R}\nonumber \\
&P(D|M_{\text{\rm Mixed}}) = w\  \frac{A_{\rm M}}{\left|\mathcal{R}\right|} \nonumber \\ &\ \ \ \ \ \ \ \ \ \ \ \ \ \ \ \ \ \  + (1-w)\  A_{\rm M}\ \mathcal{N}(x,y|\sigma_{x},\sigma_{y},\mu_{x},\mu_{y})
\end{align}
where $\mathcal{R}$ is the region bounded by the slits, $|\mathcal{R}|$ is the area of the region, $A_{\rm U/M}$ are amplitudes, and $\mathcal{N}(x,y|\sigma_{x},\sigma_{y},\mu_{x},\mu_{y})$ is a Normal distribution with parameters defined by fitting for the double diffracted X-ray signal 
\begin{align}
    \mathcal{N}(x,y\ |\ \sigma_{x},\sigma_{y},\mu_{x},\mu_{y}) = \cfrac{\exp\left(-\frac{1}{2}\left[\frac{\left(x-\mu_{x}\right)^{2}}{\sigma_{x}^{2}} + \frac{\left(y-\mu_{y}\right)}{\sigma_{y}^{2}}\right]\right)}{2\pi \sigma_{x}\sigma_{y}}.
\end{align}
Our bounds are based on a 90\% C.L. upper limit of the normally distributed contribution, $(1-w)A_{\rm M}$.
\subsection*{Data Availability}
Data displayed in this paper are available \cite{rawData}.
\bibliography{sections/bibliography.bib} %Import the bibliography file

\section*{Acknowledgements} \label{sec:acknowledgements}
This research received funding from the UK Engineering and Physical Sciences Research Council (Grants No. EP/X01133X/1 and No. EP/X010791/1). S.S., G.G., and A.A. belong to the “Quantum Sensors for the Hidden Sector” consortium funded by the UK Science \& Technology Facilities Council (Grant No. ST/T006277/1).
The experiment at SACLA was performed with the approval of Japan Synchrotron Radiation Research Institute (Proposal No. 2025B8012). The authors acknowledge the assistance of Yoshiaki Honjo, Kaneyoshi Kuwata and Kyosuke Ozaki in the preparation and operation of the CITIUS detector. C.H. acknowledges support from Trinity College Oxford’s Whitehead scholarship and general academic grant. C.H., S.Z., A.A., J.T.Y.C. and M.F. received support from the  STFC Central Laser Facility (CLF) UK Hub for the Physical Sciences on XFELs. I.I. was partially supported by Japan Society for the Promotion of Science KAKENHI (24K21199) and PRESTO (JPMJPR24J1). Assistance with sample preparation was provided through the CLF by Raj Savanam and Chris Spindloe, whom we thank for their support. A.A. acknowledges the Saudi Arabian Cultural Bureau in the UK and the Rhodes Trust for funding his research and studies.    
\section*{Author Contributions}
This project was conceived by G.G., S.S. and R.B. The experiment was designed by T.O., I.I., C.H., J.W.D.H., and G.G. carried out by C.H., J.W.D.H., T.O., I.I., S.Z., A.A., J.T.Y.C., M.F., T.H., M.N.. Data analysis was carried out by C.H. and J.W.D. H. with support from G.G., R.B. and S.S.. The manuscript was written by C.H. with support from J.W.D.H., G.G., M.N., T.O., S.S and R.B.. Further experimental and theoretical support
was provided by T.H., H.N. and A.O.T.. All authors contributed to the editing of the manuscript.
\section*{Competing interests}
The authors declare no competing interests.

% \clearpage

\end{document}